# Minimal Protocol for MRS Quality Control and Acceptance Test for Philips-Achieva MRS Tool

**Stefania Nicolosi[a], Giorgio Russo[b,c], Aldo Zucchetto[a], Giuseppe Vicari[b], Ildebrando D'Angelo[b], Domenico Messana[b], Maria Carla Gilardi[c], Paola Scifo[c]**

[a] Laboratorio Tecnologie Oncologiche (LATO), C.da Pietrapollastra, Cefalù, Italy
[b] Fondazione Istituto San Raffaele G.Gilglio (HSR-Giglio), C.da Pietrapollastra, Cefalù, Italy
[c] Istituto di Bioimmagini e Fisiologia Molecolare (IBFM-CNR), sezioni di Cefalù e Milano, via F.lli Cervi 93, Milano, Italy

**Abstract**.

Difficulties in obtaining good phantoms, improvements in technologies of voxel localization, better sequences for water and fat suppression has brought us to define a minimal Protocol of *home-made* quality controls of MRS systems.
Measurements, defined in the proposed protocol, have, as main goal, to establish if peaks quantification predicts realistic concentration values, meaning that, the occurrence of this event is a sufficient condition to declare that MRS system *works good.* Moreover, stability measurements helps in a correct data understanding. It is, indeed, realistic to think that environmental condition can introduce casual errors in the *working good* system. Discrepancies in the *working good* condition, under stochastic variability (environment), have to be related to systematic errors introduced by the set of pre and/or post-processing operations and/or by any forms of MRS *bad-working* tool that differs from the previous.
The quality control minimal protocol has been executed on a Philips-Achieva MRS system utilizing a phantom supplied by the manufacturer. The minimal protocol consists of two steps: reproducibility and performance tests. The reproducibility of the MRS measurements helps in quantifying the stability of the system. The performance test enables to establish if the system is able to reproduce concentrations in a realistic way. In both cases good results have been obtained: fluctuations of measured values are below 9% and quantification of concentration is consistent with the known values.

## 1. Introduction

Magnetic resonance spectroscopy is a powerful diagnostic tool for gaining in-vivo biochemical information and has been progressively developed over the last twenty years. MRS techniques, combined with standard magnetic resonance localization method, can now be used on clinical whole body scanners to classify and quantify the metabolites characterizing organic tissues [1]. In particular, MRS aims to distinguish among normal and pathological conditions and anables to improve non-invasively a diagnostic tool in cancer detection [2,3,4,5,6]. Good implementation of MRS acquisition procedures and post-processing analysis depends on the equipment used for the experiments. Physicists are, thus, called to perform MRS pre-in-vivo test in order to detect, resolve and prevent problems of poor-quality performance, thus enabling the highest possible standards to be maintained [7,8,9,10]. The quality tests are necessary to keep account for the in-vivo planned use of the equipment. In fact, depending on the in-vivo exams their specific acquisition sequences have to be tested in phantom study. This approach does not prevent the possibility of errors originating and, specifically, characterizing in-vivo environment but offers a good test of the selected sequences and of the quantification performance to give the possibility to exclude the fonts of errors in the successive in-vivo inspections. In this paper we propose a minimal Protocol of quality controls for MRS systems. The measurements defined in the proposed protocol have, as main goal, to establish if peaks quantification predicts realistic concentration values, meaning that, the occurrence of this event is a sufficient condition to declare that MRS system *works well.* Moreover, stability measurements help in a correct data understanding. It is, indeed, realistic to think that environmental conditions can introduce casual errors (se ci sono errori casuali il sistema non può funzionare bene). Discrepancies in the *working well* conditions, under stochastic variability (environment)(se non ci sono errori casuali=stocacisti=imprevedibili, gli errori vanno addotti al pre-processing malfatto), have to be related to systematic errors introduced by the set of pre and/or post-processing operations and/or by any forms of MRS tool *bad-working* that differs from the previous one. In the first case error can be identified and solved. The minimal measurement protocol has to be repeated in order to confirm that corrections in post and/or pre-processing operations have had a positive outcome. In the second case error has to be searched in voxel localization, in contamination of the signal by the boundary of the voxel and quality controls have to be performed in a deeper way [7,8,9,10].

## 2. Theoretical Background

In proton nuclear magnetic spectroscopy (1H-MRS) the intensity of the signal is proportional to the volume of interest and to the number of nuclei generating the signal (questo non è vero in generale, vedi spettro al carbonio). This statement, translated in frequency domain, means that the area under a specific peak is proportional to the number of nuclei (prtons, sono I protoni nucleari a precedere) precessing at that frequency. The analytical equation describing this dependence can be shown as [1]:

$$A \sim NV$$

where A indicates the area under the pick of the signal, V the volume of acquisition and N is the number of nuclei (protons) of a specific metabolite. This relationship is the theoretical basis for metabolites quantification. Unfortunately, the classification and quantification of the signal produced by MR system is a complex and rather technical issue: imperfection in acquisition and post-processing tools can introduce deviations from the theoretical behavior.

To test the linear relationship that links the number of protons with the areas of the peaks we decided to monitor the relative ratio among the areas of different known metabolites that are inside the phantom. Given two metabolites characterized by a number of protons $N_i$ and $N_{j,}$ amplitudes signal area $A_i$ and $A_j$ in the same volume of interest, the dependence between areas and concentrations implies that:

$$\frac{N_i}{N_j} = \frac{A_i}{A_j} \quad (1)$$

The knowledge of the concentrations of the phantom allows the experimental assessment of the previous equation giving us the possibility to verify the quality of our instruments. We, focusedour attention on the signals of water and acetate because they are not J coupled and therefore simpler to quantify.

The measured area depends on the choice of TE and TR. Under ideal conditions (TE = 0 and TR = ∞) the transverse magnetization decreases and the longitudinal magnetization is fully recovered. Otherwise, we must take into account the effects of relaxation and saturation. The relationship that links the parameters defined for a PRESS sequence is [1,7]:

$$A_{mis} = A_{cor}\, e^{-\frac{TE}{T_2}}\left[1 - e^{-\frac{TR}{T_1}}\right] \quad (2)$$

We named $T_1$ and $T_2$ the longitudinal and transverse decay time and $A_{cor}$ and $A_{mis}$ the correct (che è quella vera) and measured areas. substituting Ai and Aj eq. (1) becomes:

$$\frac{N_i}{N_j} = \frac{A_{mis\,1} e^{\frac{TE}{T_{2i}}} \left[1 - e^{-\frac{TR}{T_{1j}}}\right]}{A_{mis\,2} e^{\frac{TE}{T_{2j}}} \left[1 - e^{-\frac{TR}{T_{1i}}}\right]}$$

where $T_{1i}$ and $T_{1j}$ are the $T_1$ decay times of the two metabolites and $T_{2i}$ and $T_{2j}$ are their $T_2$ decay times.

From equation (2) it is easy to deduce that $T_2$ can be determined by carrying out measures of immediate $A_{mis}$ as a function of TE for a fixed value of TR. Similarly, $T_1$ can be determined by carrying out measures of $A_{mis}$ as a function of TR for a fixed value of TE (si che lo è: se tieni costante il TR la tua variabile x è il TE. Al variare di x, Amis(=la tua y) varia seguendo una legge puramente esponenziale. In particolare un esponenziale smorzato (segno – dell'esponente)). Dependence on TE is particularly easy to determine as it is the purely exponential decay. Dependence on TR is slightly more complicated and has required the empirical determination of the factor

$$A_0 = A_{cor}\, e^{-\frac{TE}{T_2}}$$

by which eq.(2) can be written in the form:

$$\frac{A_0 - A_{mis}}{A_0} = e^{-\frac{TR}{T_1}}$$

that is in the form of a simple exponential decay.

## 3. Materials

The phantom used for the measurements is a polyethylene sphere filled with a fluid of known chemical composition.

The spherical phantom contains: 1000 ml of demineralized water, 5 ml of acetate ($CH_3COOH$), 10 ml of ethanol ($CH_3\ CH_2OH$), 8 ml of phosphoric acid ($H_3PO_4$), 1 ml of Arquad solution (1%), 120 mg/ml $CuSO_4$. The proton spectrum of this phantom should have the form depicted in the figure 1

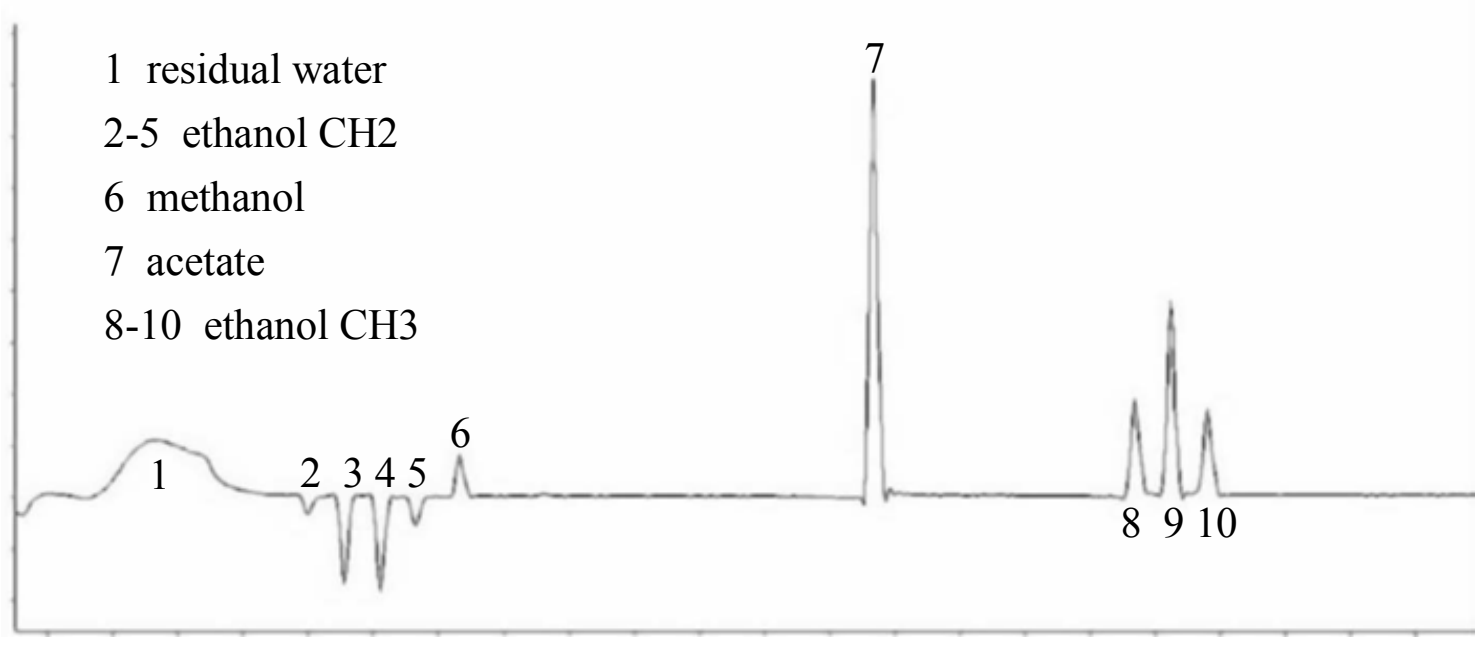

**Figura 1 - Previsional spectrum for phantom**

The signal from ethanol CH2 shows a scalar coupling with constant J~ 7Hz. The relative peak has a positive sign with a TE = 288 ms and it is inverted for TE = 144 ms as showen in the figure.

**4. Methods**

.

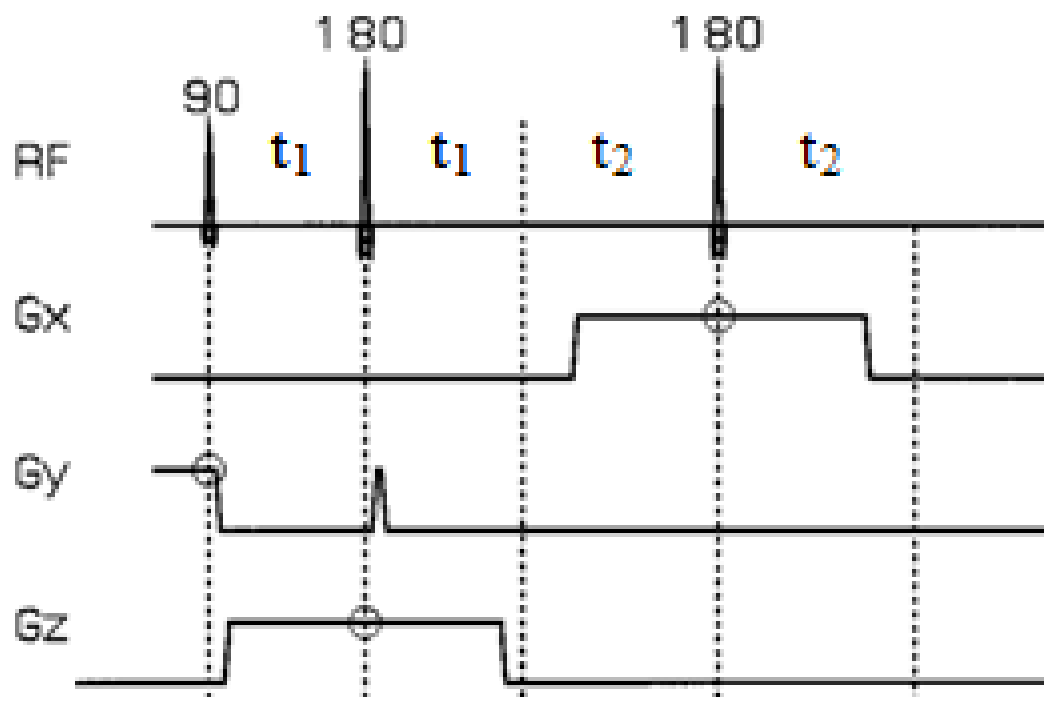

Figura 2 - PRESS

The proton spectroscopic experiments were performed on a Philips Achieva (Philips Medical system, Eindoven, The Netherlands)1.5 Tesla clinical magnetic resonance system . As in our Institution MRS will be, principally, used for the study of breast cancer, the quality control test, here described, has been executed on the same coil used in in-vivo examination: a 7 Channels Phased Array breast coil. The spectroscopic phantom was positioned inside the right hole of the coil and locked with two inert plastic compression plates to ensure immobility. The pre-scan operation has included automated shimming and (si local shimming) water suppression. [3,4,5,6,11,12] we used a single voxel PRESS sequence with water suppression, characterized by TE=288 and TR=1500, to reproduce the spectrum of the phantom. A RESS sequence has three slice-selective radio-frequency pulses with orthogonal magnetic field gradients and the intersection of the slices defines the volume of interest (VOI). The Philips post-processing software for spectroscopy enables to process the FID as prescribed by AAPM Report [7]:

a) residual water suppression,
b) phase correction,
c) zero-filling,
d) apodization filtering,
e) Fourier transformation,
f) frequency shift correction
g) baseline correction.

In order to quantify concentration, the parameter to be investigated is the area under the peak in the frequency domain. Under ideal conditions it coincides with the absorption mode. Whenever the initial phase θ of a FID is not zero, the real and imaginary channels after Fourier transform will contain mixtures of absorption mode and dispersion mode spectra:

$$Re(f) = cos\,\theta\, A(f) + sin\,\theta\, D(f)$$
$$Im(f) = cos\,\theta\, D(f) + sin\,\theta\, A(f),$$

where $Re(f), Im(f), A(f)\ e\ D(f)$ are the real, imaginary, absorption and dispersion mode of the spectrum and $f$ is the frequency. Phasing a spectrum sorts the real and imaginary channels into absorption mode and dispersion mode spectra. The Philips post-processing software supports this function in automated or manual ways. The manual phasing, being operator dependent, can introduce non-predictable errors. Therefore, we have decided to use automated phasing that, in our opinion, gives good results (la parte reale coincide con l'assorbimento e la parte immaginaria con la dispersion. Teoria delle Trasformate di Fourier.)

$[Re(f) = A(f)\ e\ Im(f) = D(f)$)].

Moreover, we have seen that the water suppression sequence combined with PRESS does not completely eliminate the water signal. Post-processing with a convolution difference filter can be used to eliminate any residual water signal. This filter applies a low-pass filter to the FID, then subtracts the filtered signal from the original data in the time domain.

Finally, spectral resolution has been improved via zero filling: the number of digital points has been increased from1024 to 2048 by adding data points of zero amplitude at the end of the signal (ok. Scrivo tutto quello che manca).

## 5. Results

4.1 *Stability Test*

During a month we have acquired and processed in this way 14 signals to establish their fluctuation and the correlated stability of the machine. The average peak area and their standard deviation for acetate $[A_{mis}\,(ACE)]$, ethanol($CH_2$) $[A_{mis}\,(ET1)]$, ethanol($CH_3$) $[A_{mis}\,(ET2)]$ and unsuppressed total water[1] $[A_{mis}\,(H_2O)]$ are:

$\overline{A_{mis}\,(ACE)} = 1{,}8 \pm 0{,}1$ $\qquad$ $\overline{A_{mis}\,(H_2O)} = 1029{,}9 \pm 50{,}6$

$\overline{A_{mis}\,(ET1)} = 1{,}7 \pm 0{,}2$ $\qquad$ $\overline{A_{mis}\,(ET2)} = 1{,}4 \pm 0{,}1$

Fluctuation rate of acetate signal is 6% , of water signal is 5%, of ethanol($CH_2$) signal is 9%, of ethanol($CH_3$) signal is 7%. These values can be consider excellent [7]. The quality control tests on long range (a tempi lunghi. È una espressione nota in letteratura.) stability will be executed weekly and one will be considered acceptable the measure having a tolerance of two standard deviations from the mean value of every metabolite (fitting. Grazie ad un programma chiamato j-miur e prodotto da un progetto Marie Curie. Aggiungerò dettagli e referenze.) .

4.2 *Quantification Performance*

As already highlighted, the area of the real part of the peak may not coincide with the absorption mode. In order to solve this problem, when we tested for stability, we adopted the technique of automatic phasing correction, but for the calculation of the decay times, in order to quantify concentrations, we have to operate in a different way. We note, first, that the dependence of the signal from damping (lo smorzamento delle diverse componenti di Fourier di uno stesso segnale: assorbimento dispersione, parte immaginaria e reale si smorzano con gli stessi tempi T1 e T2) is the same for the different components of the spectrum and does not depend on the phase. This fact implies that the module of the signal decays in the same way. Moreover, the module of the signal, by definition, does not contain information on the phase factor. Therefore, the integration of the module is equivalent to eliminate all errors related to phase factor. We stress that the quantity obtained in this way does not give information on metabolites concentration but allows us to have a more precise estimation of the decay times. The plot of the obtained experimental points is shown in figure 3 and 4 (ho ottenuto empiricamente il fattore $A_0 = A_{cor}\, e^{-\frac{TE}{T2}}$. L'avevo dichiarato nel theoretical background).

---

[1] The post processing Philips software supports a function able to resume the unsuppressed total water signal from the suppressed one so that other acquisition result unnecessary.

$T_2$ measurement: TR=1800ms

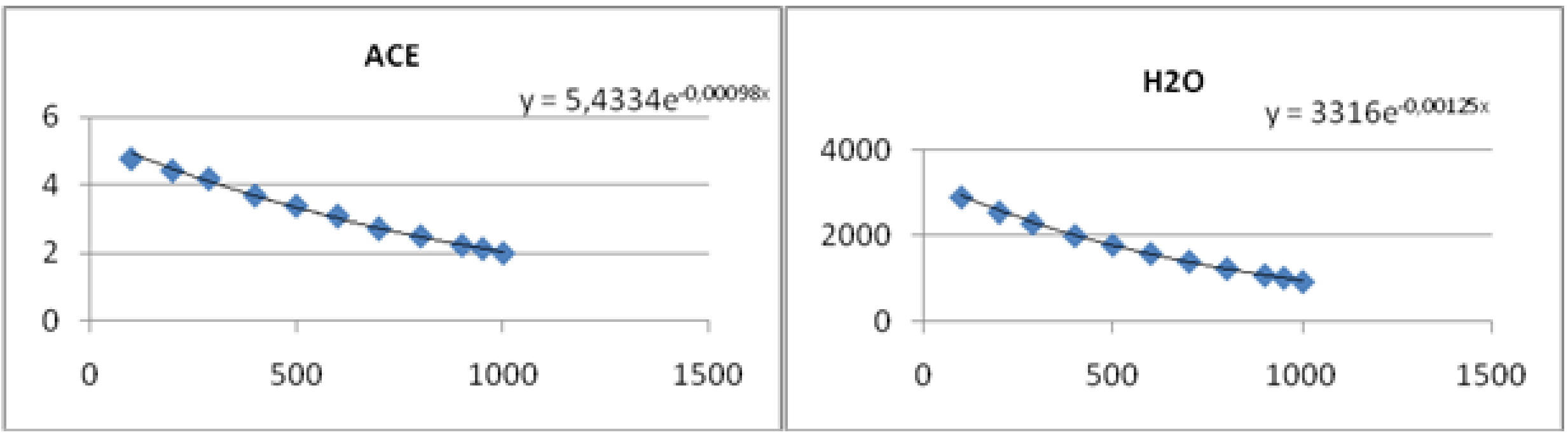

**Figure 3 - $y = A_{mis}$,x=TE, TE values are: 1000, 950, 900, 800, 700, 600, 500, 400, 300, 200, 100 ms**

$T_1$ measurement: TE=288ms

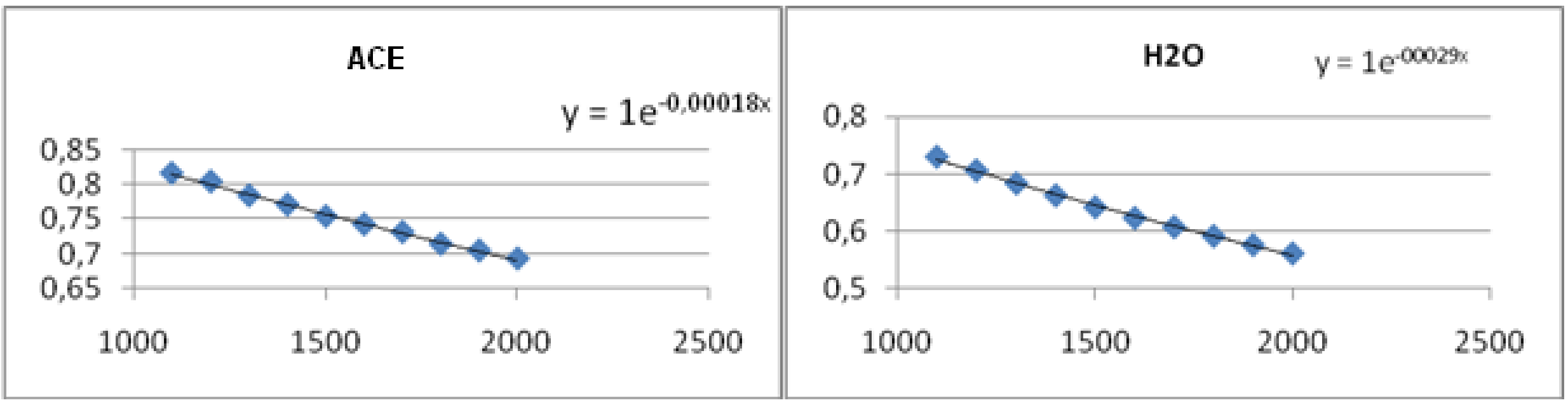

**Figura 4 - $y = \frac{A_0 - A_{mis}}{A_0}$, $x = TR$, $TR$ values are: 1100, 1200, 1300, 1400, 1500, 1600, 1700, 1800, 1900, 2000 ms**

The corresponding $T_2$ and $T_1$ values obtained by experimental data fitting are:

$T_2(ACE)$=1011,4ms $T_2(H_2O)$=795,1ms

$T_1(ACE)$=5398,9ms $T_1(H_2O)$=3429,5ms

The calculation of $T_1$ and $T_2$ allows us to find $A_{cor}$ relative to the signals of acetate and water. One must remember, however, that the analysis was developed using area data relative to the module of the signal. While, in order to find the ratio among the concentrations we need to calculate the area of the absorption components. To do this we use the average values obtained in the previous section.

Therefore, remembering that:

$A_{mis}(ACE) = 1{,}8 \pm 0{,}1$ $A_{mis}(H_2O) = 1029{,}9 \pm 50{,}6$

and that

$$\frac{N(ACE)}{N(H_2O)} = \frac{A_{corr}(ACE)}{A_{corr}(H_2O)} = \frac{A_{mis}(ACE)e^{\frac{TE}{T_2(ACE)}}\left[1 - e^{-\frac{TR}{T_1(H_2O)}}\right]}{A_{mis}(H_2O)e^{\frac{TE}{T_2(H_2O)}}\left[1 - e^{-\frac{TR}{T_1(ACE)}}\right]}$$

we obtain:

$$\frac{N(ACE)}{N(H_2O)} = \frac{A_{cor}(ACE)}{A_{cor}(H_2O)} = 0{,}0024 \pm 0{,}0002$$

This value must be compared with the ratio between the number N of protons contributing to the signal of acetate and water, respectively. Named $n(ACE)$ and $n(H_2O)$ the number of protons per molecule responsible for the signal, $m(ACE)$ and $m(H_2O)$ the molecular masses of the two compounds, $\rho(ACE)$ and $\rho(H_2O)$ their density, $V(ACE)$ and $V(H_2O)$ the volumes of the compounds in the analyzed solution, we can thus show:

$$\frac{N(ACE)}{N(H_2O)} = \frac{n(ACE)m(H_2O)\rho(ACE)V(ACE)}{n(H_2O)m(ACE)\rho(H_2O)V(H_2O)} = 0{,}0026$$

0.002625

where $n(ACE) = 3$, $n(H_2O) = 2$, $m(ACE) = 60$, $m(H_2O) = 18$, $\rho(ACE) = 1{,}05gr/cm^3$, $\rho(H_2O) = 0{,}9gr/cm^3$, $V(ACE) = 5ml$ and $V(H_2O) = 1000ml$.

The accordance among theoretical and measured value is acceptable: the theoretical value falls in the limit of the tolerance interval coupled to the measured one.

## 6. Discussion and conclusions

In the phantom study, here, the coefficient of variance of the detected intensity for signals from the Philips standard phantom were lower than 9% within a 1-month period. The $T_1$ value of water and acetate was determined to be 3429,5 and 5398,9 ms, respectively, and the $T_2$ value of water and acetate was determined to be 795,1 and 1011,4 ms, respectively. The concentration ratio of water and acetate in the standard phantom was determined to be 0,0024±0,0002, which is consistent with the value 0,0026 derived from the known volumes of metabolites provided by the manufacturer.
From the previous analysis we can consider acceptable the quality of data obtained with magnetic resonance spectroscopy exams performed on the 16 Channel, Philips Achieva, 1.5Tesla, Magnetic Resonance system here analyzed.
The quality control minimal protocol has been executed on a Philips-Achieva MRS system utilizing a phantom supplied by the manufacturer. As already highlighted, the minimal protocol consists of two steps: reproducibility and performance tests. The reproducibility of the MRS measurements helps in quantifying the stability of the system. The performance test enables to establish if the system is able to reproduce concentrations in a realistic way. In both cases good results have been obtained: fluctuations of measured values are below 9% and quantification of concentration is consistent with the known values.

The capacity to reproduce realistic values of concentrations is, as already highlighted, sufficient condition to affirm that the MRS tool coupled to the magnetic resonance image system *works well*. The *good working* condition of the system here analyzed ensures that the post and pre-processing operation have been conducted in a correct way. Other stability measurements have to be performed on our machine in order to ensure long term reproducibility of spectroscopic exams (sto testando la capacità del sistema di riprodurre valori di concentrazione= sto testando la Quantificazione, unico parametro utile in spettroscopia).

## 7. Acknowledgments

The authors would like to thank Dr. Gianni Di Leo for suggesting them some useful operation in executing pre and post-processing data analysis and the Philips local MR clinical scientist, Ing Marcello Cadioli, and MR technician, Mr Paolo Fascella to have satisfied their requests.